\documentclass[11pt]{article}
\usepackage{epsfig}
\usepackage{amsfonts}
\usepackage{rotating}
\usepackage{cite}

\def\Journal#1#2#3#4{{#1} {\bf #2}, #3 (#4)}
\def\EJPC{{\em Eur. J. Phys.} C}
\def\JPG{{\em J. Phys.} G}

\def\NPB{{\em Nucl. Phys.} B}
\def\PLB{{\em Phys. Lett.}  B}

\def\ZPC{{\em Z. Phys.} C}
\def\va{\varepsilon}

\newcommand{\Coll}{Coll.}

\newcommand{\beq}{\begin{equation}}
\newcommand{\eeq}{\end{equation}  }

\begin{document}

\hfill ANL-HEP-CP-99-99

\hfill September, 1999

\vspace{0.5cm}

\begin{center}
{\Large\bf Short-Range and Long-Range Correlations \\ \vspace{0.2cm} 
in DIS at HERA}\footnote{ 
Presented at the XXIX International Symposium on 
Multiparticle Dynamics (ISMD99), August 9-13, 1999,
Brown University, Providence, USA.}

\vspace{1.2cm}

{\large S.V. Chekanov}

\vspace{0.2cm}

Argonne National Laboratory, 9700 S.Cass Avenue, 
Argonne, IL 60439 \\E-mail: chekanov@mail.desy.de

\vspace{0.6cm}

{\large L. Zawiejski} 

\vspace{0.2cm}

Institute of Nuclear Physics, 
Kawiory 26a, PL-30-055 
Cracow, Poland \\ E-mail: leszek@mail.desy.de

\vspace{0.4cm}
For the ZEUS Collaboration 
\vspace{0.4cm}

\begin{abstract}
Correlations  in 
deep-inelastic scattering (DIS) at HERA are investigated 
in order to test perturbative QCD 
and quark fragmentation universality.
Two-particle correlations at small angular separations
are measured in the Breit frame and compared
to  $e^{+}e^{-}$ collisions.
Also presented are the correlations between the current and target
regions of the Breit frame.   
\end{abstract}

\end{center}

\section{Introduction}

Many aspects of multi-hadron production  
in DIS have extensively been  studied at HERA during  
the last years in terms of single particle 
spectra \cite{SPspectra}, 
with some attention paid to particular
aspects of correlations \cite{Corr}.
In this paper we report on recent inclusive measurements
of correlations in order to provide new and complementary
information on hadronic final state in DIS. 

There are a few important aspects concerning the correlation
studies in the Breit frame \cite{Breit} of 
DIS which should be mentioned here:

1) There exist  the analytic QCD calculations 
in the double-logarithmic approximation (DLA) 
for two-particle angular correlations \cite{Ochs}  
as well as theoretical predictions for 
current-target correlations \cite{Chek} in the 
Breit frame. The comparisons of 
these  calculations with the data     
would provide better understanding
of the analytic QCD models supplemented by 
the Local Parton-Hadron Duality (LPHD) 
hypothesis.
HERA provides a unique opportunity to confront
the analytic results, which are  usually performed at 
asymptotic energies,  
with DIS data at different $Q^2$ in order to determine 
the energy scale at which  
the calculations become unreliable.    

2) Lacking a proper theory  of  non-perturbative QCD 
necessitates the use of Monte Carlo (MC) 
models. Being beyond single-particle spectra,    
correlations provide a powerful
tool for testing these models,     
usually tuned to single-particle 
and some global observables to reproduce experimental data. 

3) In addition,
''universality`` of hadronic spectra in the current 
region of the Breit frame and
a single hemisphere of $e^+e^-$ is   
by no means an obvious observation, even for  
single-particle spectra \cite{SPspectra,H1M}. 
Correlations are more sensitive
to details in multi-hadron  production and thus can be used
to study this question in more details.   

\section{Universality and Multiplicity Distributions }

First, let us point out that deviations of the current-region 
hadronic spectra from a single hemisphere 
of $e^+e^-$ have already been observed a few years ago
by measuring multiplicity distributions  in the
Breit frame.  The results~\cite{H1M} have clearly shown that
the probabilities for events with a small number of 
tracks in the current
region are enhanced at low $Q^2$ ($Q^2\sim 10-50$ GeV$^2$). 
This effect, presumably caused
by the Boson-Gluon fusion, is less prominent at high $Q^2$.
In this region of $Q^2$, however, the initial-state gluon radiation
is significant and a similar depopulation of the current region is  
expected. This is clearly seen  for  
the multiplicity  distributions at $Q^2 > 2000$ GeV$^2$ \cite{ZEUSc}: 
The probabilities for detecting a  
small  number of charged tracks are enhanced that leads to deviations
from the standard bell-shaped form of multiplicity distributions, 
which is characteristic for  a single hemisphere of $e^+e^-$. 
For example, one of
the striking observations  is a large number of DIS events without 
particles  in the current region.  
Note that this effect is not expected
to contribute strongly  to the shapes  of  
momentum distributions in the current region, 
which are found to be similar to $e^+e^-$ \cite{SPspectra}.  

\section{Angular Correlations}

\begin{figure}[htb]
\begin{center}
\epsfig{file=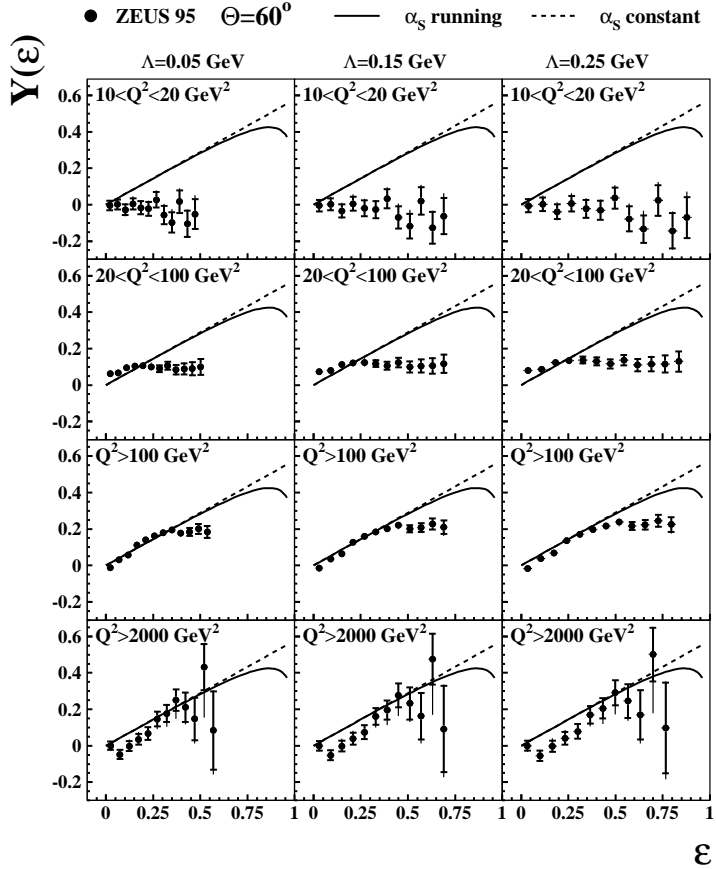,height=12.0cm}
\end{center}

\vspace*{-0.4cm}
\caption{\it Angular correlations for different $Q^2$ intervals
and values of $\Lambda$ used to  calculate $\va$. The lines
show the  DLA predictions at asymptotic energy.}
\label{fig:p2}
\end{figure}
 
Short-range correlations are sensitive to interdependence
between particles as the phase-space separations between them decrease.   
These  measurements have been 
performed in terms of angular variables \cite{ZEUSc}.     
Correlations between all charged 
particles in a cone of half-opening angle $\Theta$  defined in   
the current region of the Breit frame 
were calculated
according to the following formulae~\cite{Ochs}:
$$
Y(\va) = 
\frac {\ln(\rho(\va )/\rho_{mix}(\va ))}
{\sqrt{\ln(P\sin\Theta/\Lambda)} },  \qquad
\va =
\frac{\ln(\Theta/\theta_{12})}{\ln(P\sin\Theta/\Lambda)}, \qquad
\rho (\va ) =  \frac{1}{N_{ev}}\frac{dn_{pair}}{d\va },  
$$
in which   
the scaling variable $\va$ is used to define the 
inter-particle separation
in the inclusive two-particle density $\rho (\va )$.   
Here, $\Lambda$ is the QCD energy scale,  
$\theta_{12}$ is the relative 
angle between two particles and $P=Q/2$.   
The density $\rho_{mix}(\va )$ is calculated for 
hadrons taken from different events to remove
dynamical correlations.
Fig.~\ref{fig:p2} shows the ZEUS results together 
with the DLA predictions~\cite{Ochs} based on the LPHD hypothesis.
The correlations measured with $Y(\va )$ 
increase at high $Q^2$ as the angular
separation $\theta_{12}$ 
between  two  tracks gets smaller ($\va $ increases).   
The DLA calculation  at asymptotic energy  
fails to describe the data at low $Q^2$,  but the description    
becomes better with increase of  $Q^2$. Note that the correlations
at low $Q^2$ are hardly seen, $Y(\va )\sim 0$. This is likely due to 
small particle multiplicity in the current 
hemisphere.  

\begin{figure}[htb]
\begin{center}
\epsfig{file=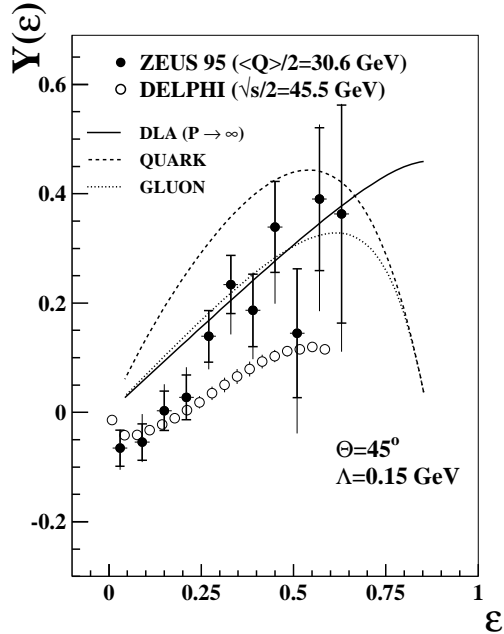,height=9.0cm}
\end{center}

\vspace*{-0.4cm}
\caption{\it Angular correlations for $Q^2>2000$ GeV$^2$
compared to $e^+e^-$ results and the DLA calculations.}
\label{fig:p2b}
\end{figure}

The ZEUS study has shown  \cite{ZEUSc} 
that the energy dependence of the correlations in Fig.~\ref{fig:p2} 
can be reproduced by MC models
which explicitly take into account 
effects related to energy momentum 
conservation in gluon splittings.   
However, there is no a single MC which can perfectly describe the data
for all  $Q^2$ regions. LEPTO and HERWIG models
show discrepancies at low $Q^2$, while ARIADNE
does not agree with the data at high $Q^2$. 
Note that a new version of ARIADNE \cite{lom} 
was  found to reproduce the data at high $Q^2$ ~\cite{ZEUSc}. 

Fig.~\ref{fig:p2b} shows the comparison between the 
ZEUS and  DELPHI
measurements \cite{cor_delphi}. Also shown  are the    
DLA calculations discussed in \cite{ZEUSc}. 
The correlations for DIS are
stronger, despite the fact that the  effective energy scale ($\sim 30.6$ GeV) 
in the current 
region of the Breit frame is smaller than that in $e^+e^-$. This
is a direct evidence for non-universality of correlations.
A possible reason for this  is kinematics of the first-order QCD
processes in the Breit frame which are  different to
those of $e^+e^-$ for one hemisphere.  

\section{Current and Target Regions}

For  the quark-parton model,
the current and target regions are well separated and 
can be viewed as completely independent
phase-space regions. The current region is 
populated by hadrons after the fragmentation
of the struck quark with $Q/2$ momentum. 
Thus this DIS region is well suited 
for any comparison of QCD calculations
for partons originating from bremsstrahlung  
of a single quark.
This situation is similar to a single hemisphere
of $e^+e^-$ in which the produced quark 
has $\sqrt{s}/2$ momentum.  Correlations  
existing between  two hemispheres of $e^+e^-$, usually called 
''forward-backward``, are due to contributions from heavy quark 
and multi-jet events. These correlations  
are small and thus can be
neglected in the theoretical description \cite{QCD}. 

\begin{figure}[htb]
\begin{center}
\epsfig{file=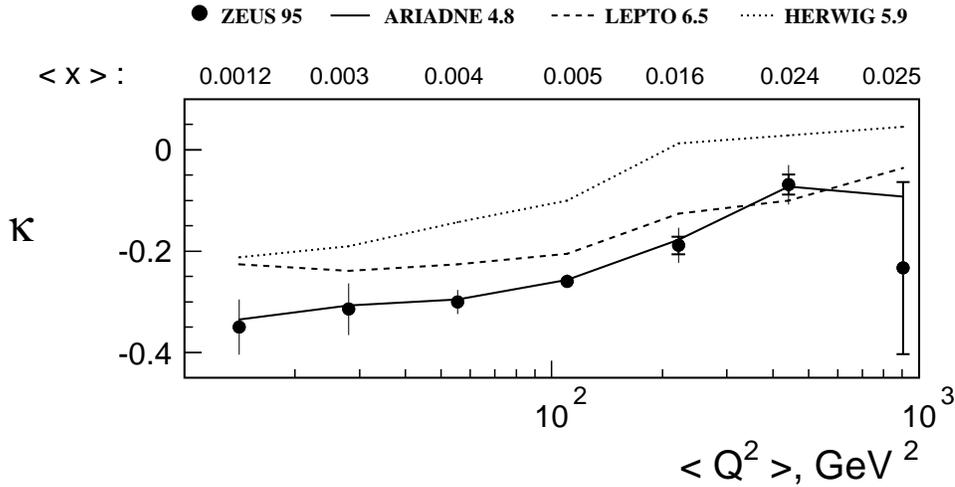,height=9.0cm}
\vspace{-0.2cm}
\end{center}
\caption{\it The evolution 
of the correlation coefficient $\kappa$ with $Q^2$.}
\label{fig:p3}
\end{figure}

In contrast to $e^+e^-$ with well separated radiations
from two initial quarks, the Breit frame does  
not provide good separation between hadrons 
associated to the final-state behavior of the proton  
and hard pQCD processes if one goes beyond the quark-parton model. 
To quantify this,
one can  look at the correlations between the total 
number of particles in the current and target regions. 
According to \cite{Chek},  
the current and target hemispheres   
cannot be treated separately when 
the first-order QCD processes are involved, such as  
Boson Gluon fusion and QCD Compton. 
In this case simple kinematic arguments 
lead to current-target anti-correlations.
The ZEUS result~\cite{ZEUSc} has confirmed such a prediction. 
The strength of this effect was 
measured with the correlation coefficient:
\beq 
\kappa = \sigma_c^{-1}\sigma_t^{-1}
(< n_cn_t > - < n_c > < n_t > ),  
\eeq 
where $n_c$ ($n_t$) is the number of particles in 
the current (target) region and
$\sigma_{c}$ ($\sigma_{t}$) is the standard 
deviation of the multiplicity distributions
in the current (target) region. The value of $\kappa$ 
is constrained by $-1\leq \kappa \leq 1$ ($\kappa = 0$ 
for  no correlations).  
Fig.~\ref{fig:p3} shows the values  of the 
correlation coefficient $\kappa$ as
a function of $Q^2$. For all $Q^2$ regions, anti-correlations are
observed ($\kappa<0$). Their magnitude   
becomes smaller as $Q^2$ increases. Note that
LEPTO and HERWIG predict smaller current-target 
interdependence, while ARIADNE agrees with the data surprisingly
well. 

\newpage 
\section{Conclusions}

For the two-particle angular correlations
significant discrepancies are observed between 
the DLA calculations and the data. This  illustrates  
non-applicability of the present analytic 
calculations for HERA energies.
The discrepancies, which  become smaller  at high $Q^2$,  
are most likely related to neglect of energy-momentum conservations
in the gluon splittings.  
Note that this effect, important for finite energies,
can be handled
in a more phenomenological approach~\cite{bp} 
to inter-particle correlations inside jets. However, the latter method 
requires a different choice for correlation observables. 

The current-target anti-correlations in DIS 
are large at low $Q^2$ and still non-negligible at rather high 
$Q^2\sim 400$ GeV$^2$. This suggests 
non-universality of hadronic final state in the current region of the
Breit frame  and a single hemisphere of $e^+e^-$ even at   
energies for which such a universality is observed for 
one-particle spectra. This phenomenon, related
to specific kinematics of the leading-order QCD processes in
the Breit frame,
can also explain the discrepancies observed 
between the ZEUS and DELPHI results for  
the angular correlations.          

{}

\end{document}